\documentclass{elsart}

\usepackage[dvips]{graphicx}
\usepackage{amssymb}
\usepackage{rotating}

\begin{document}

\begin{frontmatter}

% Title, authors and addresses

% use the thanksref command within \title, \author or \address for footnotes:
% \title{Title\thanksref{label1}}
% \thanks[label1]{}
% \author{Name\thanksref{label2}}
% \thanks[label2]{}
% \address{Address\thanksref{label3}}
% \thanks[label3]{}
% including your email address:
% \address{Address\thanksref{email}}
%\thanks[email]{E-mail: ysatou@rarfaxp.riken.go.jp}
\thanks[email]{E-mail: satou@ap.titech.ac.jp}
\thanks[present1]{Present address: 
Department of Physics, Tokyo Institute of Technology, 
Meguro, Tokyo 152-8551, Japan.}
\thanks[present2]{Present address: Soei International Patent Firm, Tokyo 
104-0031, Japan.}
\thanks[present3]{Present address: 
Department of Physics, Faculty of Engineering, 
Kyushu Institute of Technology, Kitakyushu 804-8550, Japan.} 
\thanks[present4]{Present address: 
M.\ Smoluchowski Institute of Physics, Jagiellonian University, 
PL-30059 Cracow, 
%Krak\'ow, 
Poland.} 

\title{Three-body \mbox{\boldmath $dN$} interaction 
in the analysis of the \mbox{\boldmath \nuc{12}{C}$(\pol{d},d')$} reaction 
at 270 MeV }

% use optional labels to link authors explicitly to addresses:
% \author[label1,label2]{}
% \address[label1]{}
% \address[label2]{}
\author[Riken,email,present1]{Y. Satou},
\author[Riken,present2]{S. Ishida},
\author[Tokyo_u]{H. Sakai},
\author[Saitama_u]{H. Okamura},
\author[Riken]{N. Sakamoto},
\author[Tohoku_u]{H. Otsu},
\author[Saitama_u]{T. Uesaka},
\author[Tokyo_u]{A. Tamii},
\author[RCNP]{T. Wakasa},
%\author[Riken]{K. S. Itoh},
\author[Tokyo_u]{T. Ohnishi},
\author[Tokyo_u]{K. Sekiguchi},
\author[Tokyo_u]{K. Yako},
%\author[Tokyo_u]{S. Sakoda},
\author[Saitama_u]{K. Suda},
\author[Tokyo_u]{M. Hatano},
\author[Tokyo_u]{H. Kato},
\author[Tokyo_u]{Y. Maeda},
\author[Saitama_u]{J. Nishikawa}, 
\author[Riken]{T. Ichihara}, 
\author[Kasei]{T. Niizeki}, 
\author[Bochum,present3]{H. Kamada}, 
%\author[Bochum]{A. Nogga}
\author[Bochum]{W. Gl\"{o}ckle} and 
\author[Jagiellonian,present4]{H. Wita\l a} 
\address[Riken]{Institute of Physical and Chemical Research (RIKEN), 
Wako, Saitama 351-0198, Japan}
\address[Tokyo_u]{Department of Physics, University of Tokyo, 
Bunkyo, Tokyo 113-0033, Japan}
\address[Saitama_u]{Department of Physics, Saitama University, 
Urawa, Saitama 338-8570, Japan}
\address[Tohoku_u]{Department of Physics, Tohoku University, 
Sendai, Miyagi 980-8578, Japan}
\address[RCNP]{Research Center for Nuclear Physics (RCNP), 
Ibaraki, Osaka 567-0047, Japan}
\address[Kasei]{Faculty of Home Economics, Tokyo Kasei University, 
Itabashi, Tokyo 173-8602 Japan}
\address[Bochum]{Institut f\"{u}r Theoretische Physik II, 
Ruhr-Universit\"{a}t Bochum, 
D-44780 Bochum, Germany}
\address[Jagiellonian]{Institute of Physics, Jagiellonian University, 
PL-30059 Cracow, Poland}
\begin{abstract}
% Text of abstract
We have measured the cross sections and analyzing powers $A_y$ and $A_{yy}$ 
for the elastic and inelastic scattering of deuterons 
from the $0^+$(g.s.), $2^+$(4.44 MeV), $3^-$(9.64 MeV), $1^+$(12.71 MeV), 
and $2^-$(18.3 MeV) states in \nuc{12}{C} at an incident energy of 270 MeV. 
The data are compared with microscopic distorted-wave impulse approximation
calculations 
where the projectile-nucleon effective interaction
is taken from the three-nucleon $t$-matrix 
given by rigorous Faddeev calculations 
presently available at intermediate energies. 
The agreement between theory and data compares well 
with that for the $(p,p')$ reactions at comparable incident energies/nucleon. 
\end{abstract}

\end{frontmatter}

% main text
%\section{Introduction}
Light-ion-induced inelastic scattering 
at bombarding energies above 100 MeV/nucleon 
%continues to be 
is an appealing probe of nuclear structure 
due to the simple reaction mechanism. 
In such an energy domain, 
the reaction proceeds predominantly through a single step, 
and the distorted-wave impulse approximation (DWIA) 
gives a reasonable starting point for the theoretical description of data. 
In the IA for the ($p,p'$) reaction, 
the effective interaction 
between a projectile nucleon and a target nucleon 
is taken to be the free nucleon-nucleon ($NN$) $t$-matrix. 
%and thus can be regarded as known. 
%In the case of 
For the ($d,d'$) reaction, 
the situation is 
not as simple as that for the nucleon case 
%more complicated 
because the structure of the deuteron must be considered. 
Recently Orsay group has developed 
a DWIA model~\cite{Wiele95} 
using 
%based on 
the double folding method 
to calculate the deuteron inelastic scattering at intermediate energies. 
In previous applications~\cite{Wiele95,Baker97}, 
the deuteron-nucleus ($dA$) transition matrix was calculated, 
first by folding the on-shell $NN$ $t$-matrix 
with the deuteron wave function 
to yield the deuteron-nucleon ($dN$) $t$-matrix, 
then by folding 
it 
%this projectile-nucleon effective interaction 
with the target transition density. 
In a comparison between model predictions and data, 
however, 
%it was pointed out that 
it was found that 
the $d$+$N$ elastic differential cross sections were overestimated 
by the first folding, 
%which directly leads 
leading to too large $dA$ cross sections 
by factors of 1.2--2.0~\cite{Baker97}. 
%There is a room for improvement in this respect. 
% regarding to this point. 

Present day state-of-the-art three-nucleon ($3N$) Faddeev calculations 
have made it possible 
for the $3N$ scattering processes at intermediate energies 
to be described with a reliable accuracy 
using modern $NN$ potentials~\cite{Glockle96}. 
Since the $dN$ $t$-matrix obtained from the rigorous $3N$ Faddeev calculations 
helps reduce uncertainties involved in the folding $dN$ $t$-matrix, 
it is 
quite conceivable 
%quite probable 
that the Faddeev amplitude, 
when used as an effective interaction, 
provides a more precise DWIA description of the ($d,d'$) reaction. 
Such rigorous $3N$ amplitudes have recently been successfully 
employed in a PWIA model for interpreting analyzing power data 
in the \nuc{3}{He}$(\pol{d},p)$\nuc{4}{He} reaction~\cite{Kamada00}. 
They would also facilitate analyzing deuteron spin-flip data 
taken in search for isoscalar single- and double-spin-flip 
excitations~\cite{Ysatou01}. 

This article reports on the differential cross sections 
and vector and tensor analyzing powers $A_y$ and $A_{yy}$ 
for low-lying states in \nuc{12}{C} 
excited via the $(\pol{d},d')$ reaction at $E_d$=270 MeV. 
The purpose is twofold: 
(1) provide accurate $(\pol{d},d')$ scattering data 
%for low-lying excited states 
which are scarce at intermediate energies; 
and (2) test the $3N$ amplitude given by the Faddeev calculations 
as the effective interaction in a DWIA model. 
The \nuc{12}{C} target was chosen 
as it provides both spin-flip ($\Delta S$=1) 
and non-spin-flip ($\Delta S$=0) states 
which are strongly excited via hadron inelastic scattering 
and whose structure information is available from shell-model calculations. 
Furthermore 
since the $\Delta S$=1 and $\Delta S$=0 states are excited 
from the $0^+$ ground state 
through spin-dependent and spin-independent parts 
of the effective interaction, respectively, 
we can investigate the interaction in both spin channels separately 
by using these transitions. 

%\section{Experiment}
The experiment was performed at RIKEN Accelerator Research Facility (RARF). 
The vector and tensor polarized deuteron beams of 270 MeV 
from the K=540 Ring Cyclotron were used 
to bombard a 31.3-mg/cm$^2$-thick \nuc{12}{C} target. 
Beam polarization was measured by using the $d$+$p$ elastic scattering 
at 270 MeV~\cite{Sakamoto96}. 
Typical polarizations of 60--70\% were obtained. 
The scattered deuterons were analyzed with the QQDQD-type magnetic 
spectrometer SMART~\cite{Ichihara94}. 
The angular acceptance of the spectrometer was 
100 and 50 mrad in the vertical and horizontal directions, respectively, 
%100 mr in the vertical direction and 50 mr in the horizontal direction, 
with a momentum acceptance of 4\%. 
The beam deuteron was stopped by a Faraday cup inside the scattering chamber. 
The scattering plane was perpendicular 
to the dispersive plane of the spectrometer 
due to the beam swinger system~\cite{Kato87}, 
and the scattering angle at the target 
was determined from the position at the focal plane 
normal to the dispersive plane. 
The angular resolution was less than 0.2$^{\circ}$ in FWHM, 
and the scattering angles were subdivided into 0.5$^{\circ}$ bins 
to obtain angular distributions. 
Since elastically scattered deuterons 
produced formidable count rates at forward angles, 
a movable slit was employed to stop them 
at the intermediate focusing point of SMART. 
This allowed us to take data at excitation energies as small as 1 MeV 
and at angles as small as 2.5$^{\circ}$. 
The position counter 
consisted of a 64-cm-wide and 16-cm-high multiwire drift chamber (MWDC) 
having four wire planes in both X and Y directions. 
Four plastic scintillation counters (5 mm thick) behind the MWDC 
provided pulse hight and time-of-flight information 
for particle identification. 
Fourfold coincidence of these counters generated a trigger 
for data-acquisition system~\cite{Okamura20}. 

\begin{figure}[p]
%/home/ysatou/smart/jul95f2/results/spe
\includegraphics[width=18cm,clip,angle=-90]{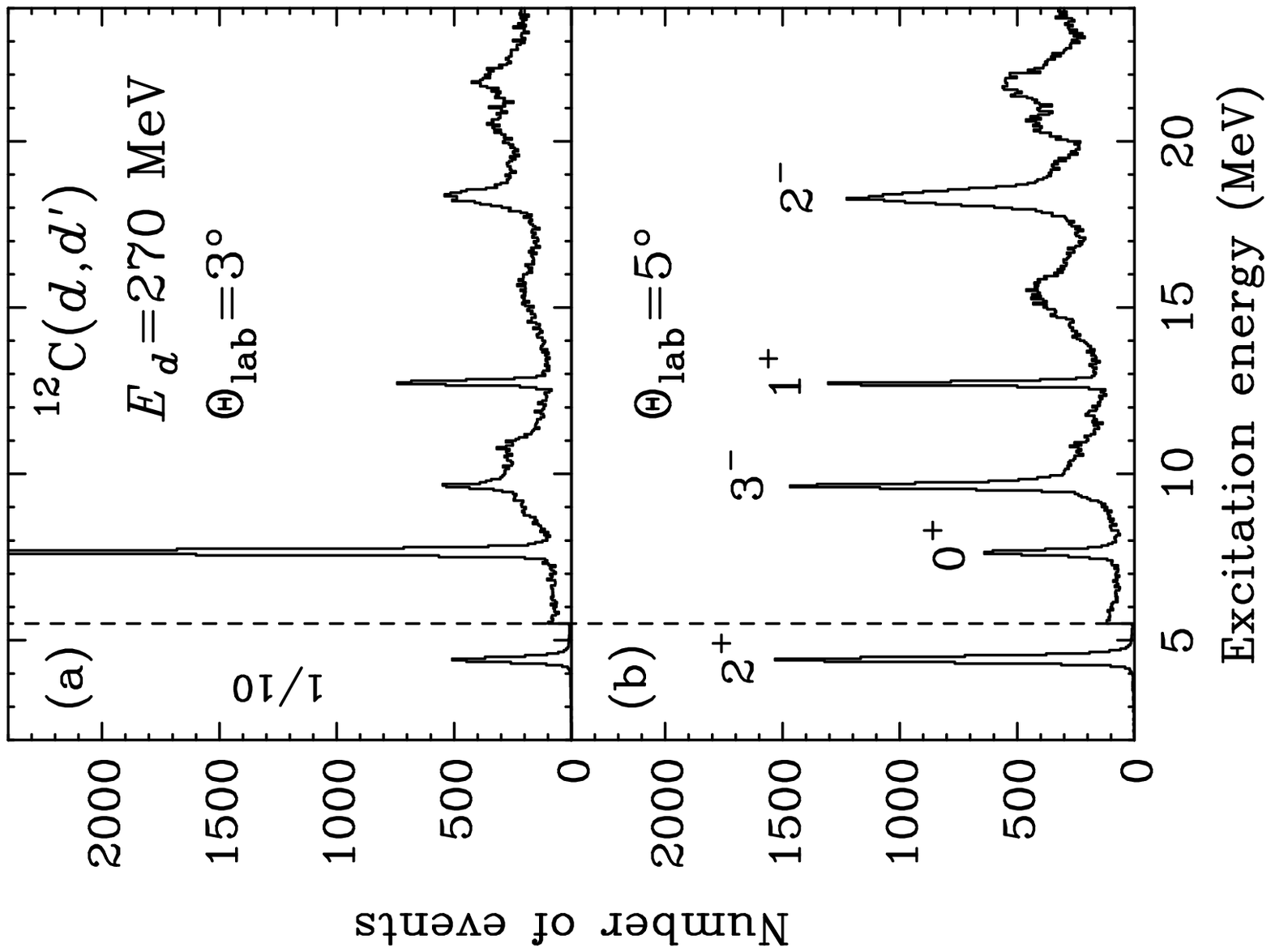}
\caption{Typical excitation energy spectra 
for the \nuc{12}{C}$(d,d')$ reaction at $E_d$=270 MeV 
at (a) ${\it \Theta}_{\mathrm{lab}}=3^{\circ}$ 
and (b) ${\it \Theta}_{\mathrm{lab}}=5^{\circ}$. }
\label{fig:12C_spectrum}
\end{figure}

%\section{Results}
Figure~\ref{fig:12C_spectrum} 
shows typical excitation energy spectra 
for the \nuc{12}{C}$(d,d')$ reaction at $E_d$=270 MeV 
at (a) ${\it \Theta}_{\mathrm{lab}}$=$3^{\circ}$ 
and (b) ${\it \Theta}_{\mathrm{lab}}$=$5^{\circ}$. 
The spin-flip $1^+$ (12.71 MeV) and $2^-$ (18.3 MeV) states 
are clearly observed 
%excited 
along with the non-spin-flip 
$2^+$ (4.44 MeV), $0^+$ (7.65 MeV), and $3^-$ (9.64 MeV) states. 
%The isoscalar $1^-$ (10.84 MeV) state is excited only weakly. 
The broad structures at 10.3 and 15.4 MeV 
are probably due to the resonances 
tentatively assigned to be $0^+$ and $2^+$ states, 
respectively~\cite{Ajzenberg90}. 
All the states are isoscalar states 
due to the isospin selectivity, 
and isovector states, 
such as the isovector $1^+$ (15.11 MeV) state, 
are entirely unseen. 
%From a comparison of relative strengths of the peaks between the two spectra, 
%some angular dependence of each peak can be inferred. 

\begin{figure}[p]
%/home/ysatou/smart/jul95f2/results/spe
\includegraphics[width=14cm,clip]{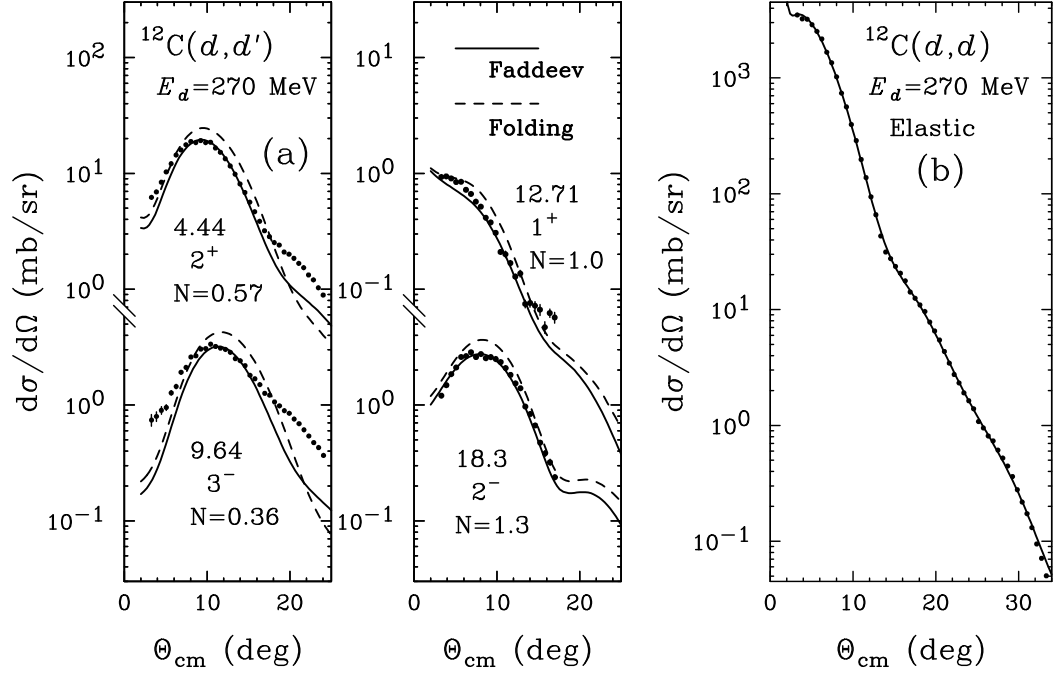}
\caption{(a) Measured differential cross sections 
for the \nuc{12}{C}$(d,d')$ reaction at $E_d$=270 MeV 
leading to low-lying excited states in \nuc{12}{C} 
are shown as full circles. 
The solid (dashed) lines are results of the DWIA calculations 
using the projectile-nucleon effective interaction 
derived from the Faddeev (folding) calculations. 
(b) Measured differential cross sections 
for the elastic scattering of deuterons from \nuc{12}{C} at $E_d$=270 MeV. 
The solid line shows the result of the optical model calculation 
using the parameters listed in Table.~\ref{tbl:optical_parameters_12C}. }
\label{fig:dsd}
\end{figure}

\begin{figure}[p]
%/home/ysatou/smart/jul95f2/results/spe
\includegraphics[width=14cm,clip]{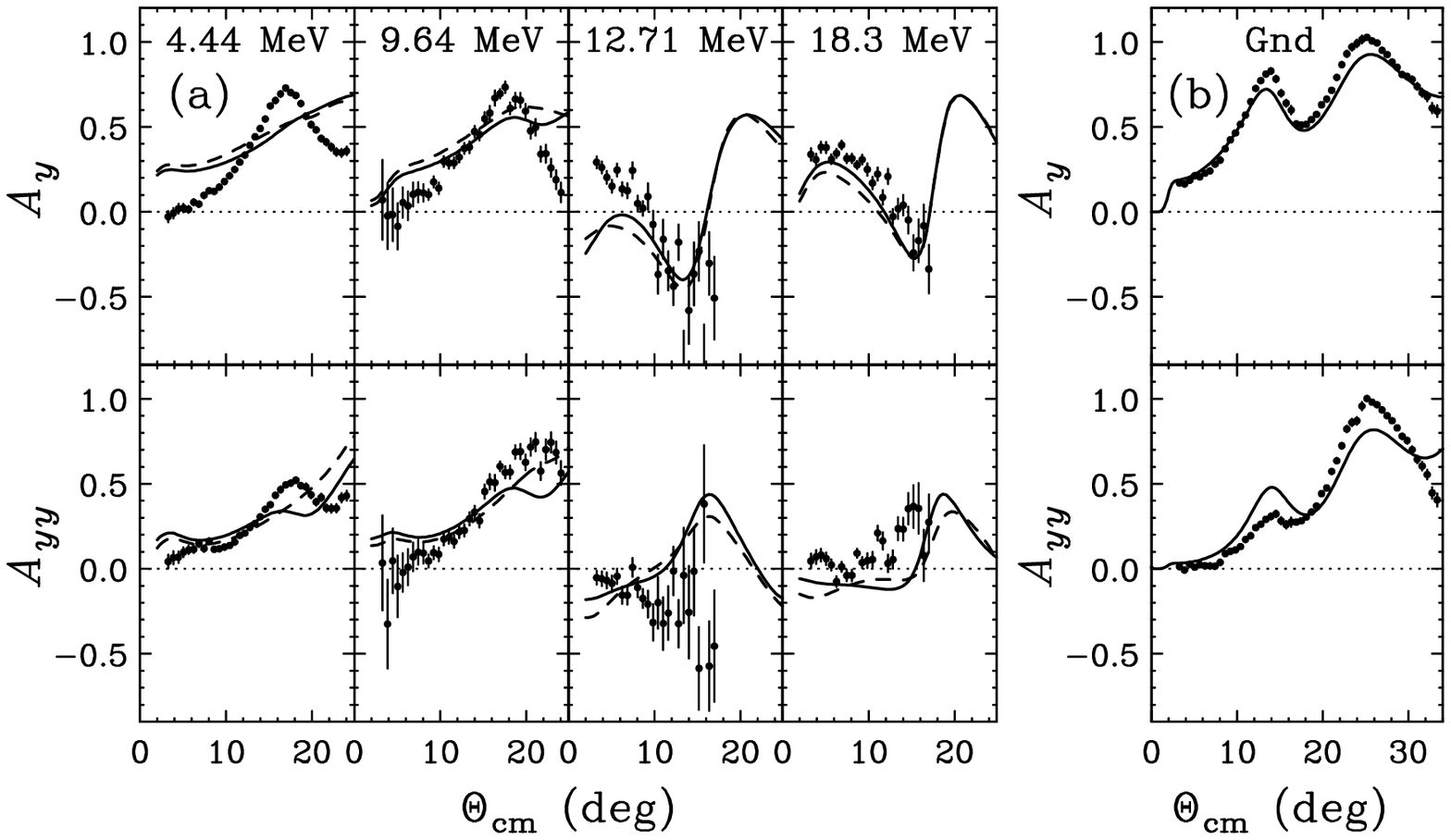}
\caption{Same as Fig.~\ref{fig:dsd}, 
but for vector and tensor analyzing powers $A_{y}$ and $A_{yy}$. }
\label{fig:azp}
\end{figure}

The spectra were analyzed 
by using a peak fitting program {\sc specfit}~\cite{Blok75} 
to extract yields contained in each peak. 
The cross sections and analyzing powers $A_y$ and $A_{yy}$ 
were calculated from the yields, 
and the continuum background and other overlapping states 
were subtracted from the data. 
%From the yields 
%cross sections and analyzing powers $A_y$ and $A_{yy}$ were calculated. 
%The continuum background and other overlapping states 
%are thus subtracted from data. 
%Figures~\ref{fig:dsd}(a) and \ref{fig:azp}(a) 
%show respectively the cross sections and analyzing powers 
%for the $2^+$, $3^-$, $1^+$ and $2^-$ states. 
The experimental cross sections and analyzing powers 
for the $2^+$, $3^-$, $1^+$, and $2^-$ states 
are shown as full circles in Figs.~\ref{fig:dsd}(a) 
and \ref{fig:azp}(a), respectively. 
The error bar includes 
the statistical error and the error from the fitting procedure. 
The systematic uncertainty in the absolute magnitude of the cross section 
is estimated to be less than 10\% 
taking account of ambiguities 
in charge collection, target thickness and solid angle. 
The systematic uncertainties for $A_y$ and $A_{yy}$ 
are 2\% and 6\%, respectively, 
which come from the normalization of beam polarizations. 
The excitations of the $1^+$, $2^-$, $2^+$, and $3^-$ states 
are dominated by transitions with transferred angular momenta $\Delta L$ 
of 0, 1, 2, and 3, respectively. 
Cross section data show angular distributions 
which are characteristic of $\Delta L$. 
In contrast, 
analyzing power data depend on $\Delta L$ only weakly, 
while they are characterized by transferred spin values $\Delta S$. 
For example, $A_y$ monotonically increases 
for the non-spin-flip $2^+$ and $3^-$ states 
for an angular range between 3$^{\circ}$ and 15$^{\circ}$, 
while it decreases in the same range 
for the spin-flip $1^+$ and $2^-$ states. 
Such a unique $\Delta S$-dependence of the analyzing powers 
%at forward angles in the $(\pol{d},d')$ reaction 
can be used as a signal of spin transfer 
for a given state under investigation. 

%for states with unknown spin and parity. 

Measured cross sections and analyzing powers 
of the \nuc{12}{C}$(d,d)$ elastic scattering at $E_d$=270 MeV 
are shown as full circles in Figs.~\ref{fig:dsd}(b) 
and \ref{fig:azp}(b), respectively. 
The optical potential parameters were determined by fitting the data 
using the code {\sc ecis}~\cite{Rayn72}. 
The results of the optical model fit are shown as solid lines 
in Figs.~\ref{fig:dsd}(b) 
and \ref{fig:azp}(b). 
The deduced parameters are listed in Table~\ref{tbl:optical_parameters_12C}. 
They are consistent with the systematics of parameters 
at different energies~\cite{Wiele95,Wiele94}. 

\begin{sidewaystable}[h]
\caption{Optical potential parameters 
obtained from the analysis of the present elastic scattering data 
of deuterons from \nuc{12}{C} at $E_d$=270 MeV. 
The potential is given by 
$U(r)  =V_Rf(x_R)+{\mathrm i}W_If(x_I) 
      -\left[ 
       V_{RSO}\frac{1}{r}\frac{\d}{\d r}f(x_{RSO}) 
      +{\mathrm i}W_{ISO}\frac{1}{r}\frac{\d}{\d r}f(x_{ISO})\right] 
       \mbox{\boldmath $L$}\cdot\mbox{\boldmath $s$} 
      +V_{\mathrm{Coul}}(r_C), \label{Optical_potential} $
where $f(x_i)=\left[1+\exp(x_i)\right]^{-1}$ 
with $x_i=(r-r_iA^{\frac13})/a_i$. 
The Coulomb radius parameter is $r_C$=1.3 fm. }
\label{tbl:optical_parameters_12C}
\vspace{3mm}
\begin{tabular}{cccccccccccc} \\ \hline
 $V_R$   & $r_R$    & $a_R$   & $W_I$   & $r_I$   & $a_I$    & 
$V_{RSO}$& $r_{RSO}$&$a_{RSO}$&$W_{ISO}$&$r_{ISO}$&$a_{ISO}$ \\
 (MeV)   & (fm)   & (fm)   & (MeV)   & (fm)   & (fm)         & 
 (MeV)   & (fm)   & (fm)   & (MeV)   & (fm)   & (fm)         \\ \hline
$-$19.27 & 1.41 & 0.75 &$-$19.64 & 1.08 & 0.89 & 
$-$7.20  & 0.91 & 0.71 &    1.64 & 0.89 & 0.71 \\ 
\hline
\end{tabular}
\end{sidewaystable}

%\section{DWIA}
\label{DWIA}
Microscopic DWIA calculations were performed 
using the formalism described in Ref.~\cite{Wiele95}. 
The $T$-matrix 
in the $dA$ system 
is given by 
\begin{eqnarray}
T^{\mathrm{DWIA}}_{dA}=\langle 
X^{(-)}\chi_{d'}{\it \Phi}_{A^*}|t_{dN}
\e^{\mathrm{i}\vec{q}\cdot(\vec{R'}-\vec{R})}|
X^{(+)}\chi_{d} {\it \Phi}_{A} \rangle , \nonumber
\end{eqnarray}
where the distorted waves in the initial and final channels 
are denoted by $X^{(+)}({\vec R})$ and $X^{(-)}({\vec R})$, 
the target wave functions by 
${\it \Phi}_{A}(\vec{R'})$ and ${\it \Phi}_{A^*}(\vec{R'})$, 
and the deuteron spinors by $\chi_{d}$ and $\chi_{d'}$, respectively. 
${\vec q}={\vec k}_{\mathrm{in}}-{\vec k}_{\mathrm{out}}$ 
is the momentum transfer, 
where ${\vec k}_{\mathrm{in}}$ and ${\vec k}_{\mathrm{out}}$ 
are the incoming and outgoing deuteron momenta, respectively. 
The on-shell $dN$ $t$-matrix $t_{dN}$ is used 
as the projectile-nucleon effective interaction. 
In the $dN$ c.m.\ frame this is given by~\cite{Tsuzuki94} 
\begin{eqnarray}
t_{dN}(q)=\alpha 
 &+&\beta S_n+\gamma \sigma_n+\delta S_n \sigma_n +\epsilon S_q \sigma_q 
  + \zeta S_p \sigma_p +\eta Q_{qq} +\xi Q_{pp} \nonumber \\
 &+&\kappa Q_{qq} \sigma_n +\lambda Q_{pp} \sigma_n +\mu Q_{nq} \sigma_q  
  + \nu Q_{np} \sigma_p, \nonumber
\label{eqn:tdn}
\end{eqnarray} 
where $\sigma$ is Pauli spin matrix, $S$ and $Q$ deuteron spin operators, 
and 
the coefficients $\alpha$ through $\nu$ 
are complex parameters which depend on the incident energy and $q$. 
Unit vectors are given by 
$\hat{q}$, 
$\hat{n}//{\vec k}_{\mathrm{in}}\times {\vec k}_{\mathrm{out}}$
and 
$\hat{p}=\hat{n}\times \hat{q}$. 
We examined two different $dN$ interactions: 
(1) the Faddeev interaction 
given by the $3N$ Faddeev calculations at $E_d$=270 MeV, 
in which the total angular momenta of the two nucleon system up to $j$=5 
are taken into account~\cite{Glockle96}; 
and (2) the folding interaction 
obtained by folding the on-shell $NN$ $t$-matrix 
at half the incident deuteron energy 
with a full deuteron wave function~\cite{Wiele95}. 
In both calculations the CD-Bonn potential~\cite{Machleidt94} was used 
for the $NN$ interaction and for the deuteron wave function. 
% In reply to comment 2-a of Referee A the following one sentence 
% has been changed and one sentence has been added. 
A good agreement has been found between 
the predictions of the Faddeev calculations 
and the $\vec{d}$+$p$ elastic scattering data 
at 270 MeV~\cite{Sakai00,Sekiguchi02}. 
In contrast, 
only a fair agreement could be obtained 
with the folding calculations~\cite{Ysatou01}; 
for instance the calculated cross section is about 1.5 times larger 
than the experimental one at ${\it \Theta}_{\mathrm{c.m.}}$=50$^{\circ}$ 
where the Faddeev result almost coincides with the data. 
% Modifications were made up to here. 
The distorted waves $X$ 
%in the entrance and exit channels 
were generated by using the optical potential parameters 
in Table~\ref{tbl:optical_parameters_12C}. 
The target wave functions ${\it \Phi}$ 
were those of Cohen and Kurath~\cite{Cohen65} 
and Millener and Kurath~\cite{Millener75} 
for positive and negative parity states, 
respectively. 
To account for the effect of core polarization 
the spectroscopic amplitudes for natural parity states 
were renormalized to reproduce the observed electric transition 
probabilities~\cite{Ajzenberg90}. 
The single particle wave functions were those of a harmonic oscillator well, 
with the center of mass motion corrected in $q$-space 
following the Ref.~\cite{Tassie58}. 
The integral over $q$ in $T^{\mathrm{DWIA}}_{dA}$ was carried out 
over the range of $q$ where $t_{dN}$ is known: 
$q_{\mathrm{max}}$=3.4 and 2.5 fm$^{-1}$ 
for the Faddeev and folding interactions, respectively. 
%($q_{\mathrm{max}}$=3.4 (2.5) fm$^{-1}$ 
%for the Faddeev (folding) interaction). 
Since the form factors decreased rapidly with $q$ for states examined, 
the results with the Faddeev interaction 
did not depend sensitively on the choice of the $q_{\mathrm{max}}$ values. 

%\section{Discussion}
The calculated cross sections using the Faddeev and folding interactions 
are respectively 
shown as solid and dashed lines in Fig.~\ref{fig:dsd}(a). 
The curves are normalized with values indicated in the figure. 
For natural parity transitions normalization factors of around 0.5 
are required, 
while for unnatural parity transitions the factors are around unity. 
The theoretical cross sections obtained with the folding interaction 
overestimate those with the Faddeev interaction by about 30\% near the peak 
for both $\Delta S$=0 and $\Delta S$=1 transitions. 
% In reply to comment 2-b of Referee A the following one sentence 
% has been changed. 
The difference between the two curves 
is ascribed to higher order processes within the projectile-nucleon system, 
such as the multiple scattering, virtual break-up and/or rearrangement, 
which are included in the Faddeev interaction but not in the folding one. 
% Modifications were made up to here. 
% In reply to comment 3 of Referee A the following two sentences 
% has been changed, resulting in three sentences. 
%The problem associated with the folding interaction 
%(the cross section overestimation) 
%was previously identified only in the $\Delta S$=0 channel. 
%The present result suggests 
%that there exists a similar problem in the $\Delta S$=1 channel as well. 
It is to be noted that such an effect of correlation 
among the interacting three nucleons (cross section reduction near the peak), 
previously noted on the basis of the comparison 
of the $d$+$N$ elastic cross sections  
with the folding model calculations~\cite{Baker97}, 
has been primarily concerned in the $\Delta S$=0 channel. 
This is because the $d$+$N$ elastic amplitude 
is dominated by the isoscalar spin independent ($\Delta S$=0) part 
of the effective interaction, 
especially at low momentum transfer region 
where the cross section reaches the maximum. 
Therefore the present results suggest that 
there clearly exists a similar effect of correlation, 
to reduce cross sections, in the $\Delta S$=1 channel as well. 
% Modifications were made up to here. 
The shapes near the peak in the angular distribution 
are well reproduced 
with a harmonic oscillator size parameter $b$=1.76 fm 
determined from elastic electron scattering on \nuc{12}{C}~\cite{Jones94}, 
except for the $3^-$ state for which a larger value of $b$=1.90 fm 
is required. 
Such a larger value of $b$ for the $3^-$ state 
is consistent with 
the analysis of the $(p,p')$ reaction~\cite{Comfort82}. 
%the analyses of the $(e,e')$ and $(p,p')$ reactions~\cite{Comfort82}. 
%and may indicate that the wave function for this state has a larger size 
%than the other states considered here. 

Calculated analyzing powers with the Faddeev and folding interactions 
are shown in Fig.~\ref{fig:azp}(b) as solid and dashed lines, respectively. 
The Faddeev interaction 
gives 
%gave 
results 
which differ only in details from those given by the folding interaction. 
Both calculations reproduce qualitative features of the data. 
However, neither of them gives a full description 
of the detailed oscillating patterns of analyzing powers 
for natural parity states, 
and of the forward angle behavior of $A_y$ for the $1^+$ state 
where the data show positive values 
while the calculations exhibit negative values. 
Such discrepancies may result from processes not treated by the present DWIA, 
such as those arising from the presence of nuclear medium 
where the struck nucleon is embedded. 
The treatment of the deuteron as a single unit during the distortion process 
may also be responsible for the failure of the calculation. 

The normalization factors for the calculated cross sections 
of around 0.5 required for natural parity states 
are consistent with those found in $(p,p')$ studies 
in the 100--200 MeV range~\cite{Comfort82,Comfort81}. 
% In reply to comment 1 of Referee A two sentences were added. 
The factors, however, are different from the ones in electron scattering, 
which gives the normalizations close to unity~\cite{Flanz78}. 
It is likely that the use of a density-dependent interaction~\cite{Kelly80} 
and/or a fully microscopic optical potential~\cite{Hicks88,Dortmans95} 
helps solve the normalization problem. 
% Modifications were made up to here. 
The normalization factor of unity for the $T$=0 $1^+$ state 
is consistent with that obtained 
by Willis {\it et al.}~\cite{Willis91} at $E_p$=200 MeV, 
who used the $NN$ $t$-matrix in $q$-space 
directly as the projectile-nucleon effective interaction, 
similarly to the present analysis. 
In other $(p,p')$ analyses for the $T$=0 $1^+$ state in the same energy range 
the Love and Franey interaction~\cite{Love81} was employed 
for the effective interaction. 
%as the projectile-nucleon effective interaction. 
%In the other $(p,p')$ analyses for the $T$=0 $1^+$ state 
%in the same energy range 
%the Love and Franey effective interaction~\cite{Love81} was utilized. 
It was found that the calculated cross sections 
%consistently 
overestimated the data by a factor of 2 
at forward angles~\cite{Comfort82,Comfort81,Bauhoff83}. 
From the studies, however, 
the origin of the discrepancy could be identified 
neither in terms of the nuclear structure nor the effective interaction. 
%It was found that the calculated cross sections 
%consistently overestimated the data at forward angles by a factor of 2, 
%the origin of the discrepancy being identified 
%neither in terms of the nuclear structure 
%nor the effective interaction~\cite{Comfort82,Comfort81,Bauhoff83}. 
Note that inelastic electron scattering 
is insensitive to the $\Delta S$=1 isoscalar transitions, 
and the nuclear structure for the $T$=0 $1^+$ state at 12.71 MeV 
is not as well understood as that for the $T$=1 $1^+$ state at 15.11 MeV. 
Present analysis with the Faddeev interaction, 
giving a satisfactory description for the cross section 
in both magnitude and shape, 
suggests that there is little reason 
that the spectroscopic terms of Cohen and Kurath~\cite{Cohen65} 
for this $1^+$ (12.71 MeV) transition contain errors. 
In the \nuc{12}{C}$(d,d')$ study at $E_d$=400 MeV, 
it was pointed out 
that the DWIA cross sections using the folding interaction 
for the $1^+$ (12.71 MeV) state 
were larger than the data by factors of 1.5--2.0 
at forward angles~\cite{Wiele95,Johnson95}. 
It is interesting to see if the use of the Faddeev interaction at 400 MeV 
could reduce the DWIA cross sections 
so that the calculated cross sections might 
in fact come close to the experiment. 

%\section{Summary}
In summary, 
we have measured the cross sections and analyzing powers 
for low-lying states in \nuc{12}{C} 
by the $(\pol{d},d')$ reaction at $E_d$=270 MeV. 
Microscopic DWIA calculations were performed 
by using the three-nucleon ($3N$) $t$-matrix given 
by rigorous Faddeev calculations 
as the effective interaction. 
All the characteristic features of the data are reproduced satisfactorily. 
%A good overall agreement was obtained between the data and the calculations. 
Normalization factors required for calculated cross sections 
are consistent with those obtained from comparable analyses 
of the $(p,p')$ reactions at similar incident energies. 
Correlations among nucleons in the projectile-nucleon system 
are found to reduce peak cross sections by about 30\% 
in both $\Delta S$=0 and $\Delta S$=1 channels. 
This work represents the first application 
of the $3N$ Faddeev amplitude presently available at intermediate energies 
in a DWIA analysis of the $(d,d')$ reaction as an effective interaction. 
Such a rigorous $3N$ amplitude will find a wide range of new applications 
for intermediate energy nuclear spectroscopy. 

%The use of rigorous $3N$ Faddeev amplitudes in the study 
%of light-ion-induced scattering at intermediate energies 
%is in its early stage, 
%but one may expect such applications to grow. 

%\section{Acknowledgments}
We are grateful to the staff of RARF, 
particularly, Dr.\ Y.\ Yano, Dr.\ A.\ Goto and Dr.\ M.\ Kase 
for their invaluable assistance through the experiment. 
This work was supported financially in part 
by the Grant-in-Aid for Scientific Research No.\ 04402004 
of Ministry of Education Science and Culture of Japan, 
and the Special Postdoctoral Researchers Program 
at the Institute of Physical and Chemical Research (RIKEN). 

\end{document}